# Utilizing Low-Cost Linux Micro-Computer & Android Phone Solutions on Cube-Satellites

Ahmed Farid[1], Ahmed Samy[2], Ahmed Shalaby[2], Ahmed Tarek[2], Mahmoud Ayyad[2], Muhammad Assem[2], Samy Amin[2]

[1] B.Sc. Graduation project, Computer Engineering Department, October University for Modern Sciences & Arts

[2] B.Sc. Graduation project, Aerospace Engineering Department, Cairo University

Realizing functional space systems using flight-tested components is problematic in developing economies, as such components are costly for most institutions to sponsor. The B.Sc. project, Subsystems for 2$^{nd}$ Iteration Cairo University Cube-Satellite, addresses technology demonstration using commercially available electronics and low cost computing platforms, such as Android phones and Raspberry Pi Linux micro-computer as computing hardware. As for software, the project makes use of open-source modules and locally developed code to implement needed functionalities, in addition to a mechanism to operate a virtual desktop Linux OS in parallel to an Android application. The paper aims to demonstrate the significance, operation design, and problem solving of such approaches. The paper concludes with future prospects for improving upon the proposed computing systems.

## 1. Introduction

Onboard computing systems act as the backbone of any satellite, governing its operations and data transactions amongst subsystems, with ground-stations, and other satellites. According to a self-initiated survey, there were two commonly used hardware strategies for onboard computing: Micro-controllers (Example: PIC16f877), and ready-made flight-tested computers [1,2]. For the micro-controllers strategy, components are available and they are useful for prototyping functionalities, but building a finalized integrated system requires time and meticulous from-the-ground-up design, which eventually adds complexity. Ready-made computers ease the development and prototyping phases and are reliable in space environments. However, the downside is that they are expensive and not easy to sponsor generally in developing economies.

One of the aims of the B.Sc. project, Subsystems for 2$^{nd}$ Iteration Cairo University Cube-Satellite, is to overcome such downsides by implementing the computing system using locally available low-cost hardware. The project implemented two computing systems based on two different hardware strategies: Low-cost Linux micro-computer (Example: Raspberry Pi, which will be explained), or an Android smartphone (As recently done by NASA and SSTL). The paper shows the significance of such low-cost utilizations that play a key-role in capacity-building, problem solving, and joining the space race for growing economies.

## 2. Defining specifications

It is required to implement an onboard computing system for a cube-satellite within circumstances of local unavailability and lesser funding. To further define the specifications, the general requirements are as follows:

- Autonomous operation.
- Operating a payload camera
- Providing necessary interfaces. (i.e. I2C, SPI, A/D… etc.)
- Sensor readings acquisition.
- Support for attitude determination and control mechanisms.
- Real-time system monitoring.
- Bidirectional communications handling.

In lights of such requirements, the system is needed to support fast and real-time processing, memory space for code and stored camera and log files, and low-level hardware IO mechanisms. Taking the lesser funding and local unavailability into account, Linux micro-computers and Android smartphones are perfect candidates for developing such system.

## 3. Linux Micro-Computers

Recently, small-sized computer chips emerged into the consumer market with main intentions of teaching the youth about computing, or for engineering projects. These computer chips come with processors ranging between 500-1000 MHz speed, 512 Mb - 1Gb RAM, USB, Ethernet, and the ability to host Linux as an operating system. The most important trait of them all is the ultra-low price tag as they start from as low as 25$. Examples are the Raspberry Pi, BeagleBoard, ODROID, and many more [3].

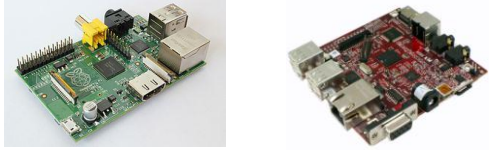

Fig. 1. Raspberry Pi and BeagleBoard XM.

The main significance in using such micro-computers lies in their ultra low-cost, high processing power with respect to flight-qualified computers, and sizes small enough to fit in a 1U cube-satellite. Another point of significance lies in the ability to host Linux as an operating system, thus giving the developers the power to make use various programming platforms and applications originally targeted the desktop on the level of embedded systems (As to be explained). It should also be noted that these computers support low-level electronic interfaces (Digital IO, UART, $I^2C$… etc.), and removes the need to fully depend on a USB-to-TTL converter. At the time of development, the B.Sc. project used the Raspberry Pi [4] for the micro-computer strategy due to its availability, acceptable capabilities, and price tag.

As stated, the project makes use of a Raspberry Pi as an onboard computer for the cube-satellite. The specifications are as follows:

Table 1. Raspberry Pi specifications

| Processor | ARM11 CPU, 700 MHz |
|---|---|
| Memory | 512 Mb of RAM<br>Up to 32 Gb SD Card storage |
| Size | 85.60 x 54 x 18 mm |
| Weight | 45 gm |
| Power | 3.5 W: 5 v – 700 mA (350 mA without display output) |
| Interfaces | USB<br>Ethernet<br>3.5 mm Audio output<br>HDMI/VGA<br>Hardware IO: Digital, UART, SPI, $I^2C$   (All are 3.3 v) |

The onboard computer is naturally required to interact with other subsystems. The paper discusses the following:

- Computation and operations management
- Sensor readings acquisition and actuator control.
- Obtain and store images from a camera.
- Receive/Transmit data using a transceiver.

### 3.1. Computation and operations management

Any onboard computer's task is to perform computations based on inputs and control the flow of operations. Since the computer operates on Linux, the freedom of using various programming languages for the subroutine management application is available. For the project, Python is utilized as it is highly supported on the Raspberry Pi, and allows writing minimal, yet functional code. In any case, the management software is considered a main thread that receives inputs, and initiates subroutines as other threads accordingly. The conditions and states are determined according to the satellite's mission; state computation generally can be triggered by GPS locations, incoming ground-station commands, sensor readings, end of a subroutine, or as the mission requires.

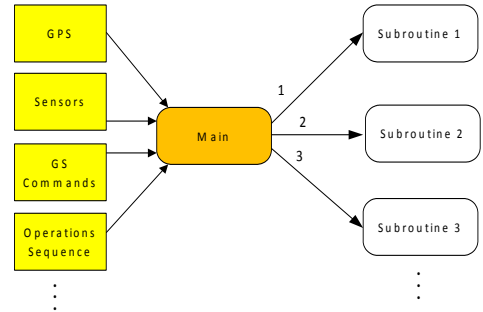

Fig. 2. Main function starts a subroutine based on a computed state.

In addition to the management software, the computer is required to do attitude determination computations for the satellite's actuator system, based on sensor readings. This was implemented in the project, but as a Matlab script. In order to minimize code porting time and to make use of Linux, the open-source mathematical environment Octave [5] is installed and used to run these Matlab scripts locally on the micro-computer on the shell. This gives the opportunity of including Matlab-style computations not only to light-weight environments as the Raspberry Pi, but also to embedded systems generally.

### 3.2. Sensor readings acquisition and actuator control

The use of a Raspberry Pi mainly has a drawback of hardware IO limitation [6]; it does not have an onboard A/D converter for analog readings or PWM for analog output. In addition, there are only 17 pins suited for programming, but some of which are reserved for certain interfaces. The drawback is more problematic as there is a local unavailability of some sensors for $I^2C$ or SPI interfaces.

Fig. 3. Raspberry Pi pinout. No analog input or PWM included.

Despite such drawback, the hardware limitation is solved by a simple interface with an external micro-controller that supports the missing needed interfaces and addition IO pins. In fact, this solution technically cuts down time needed to port publicly available code and libraries of sensor interfaces used on micro-controllers to the micro-computer. To add even more value, utilizing Atmel Atmega328 micro-controllers give the opportunity to make use of public and open-source codes and libraries targeting the Arduino platform [7]. This proved particularly useful as it eased the prototyping process and reduced development time.

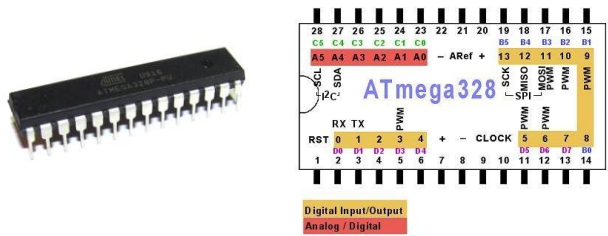

Fig. 4.  Atmega328 pinout.

There are two options for such interface:
- I$^2$C bus interface
- UART serial link

The I2C bus strategy is useful as it creates a single two-line bus where all components are connected. This is also suitable when most of the components are I2C capable. However, the downside lies in the managing of addressing within the bus, which adds complexity [8].

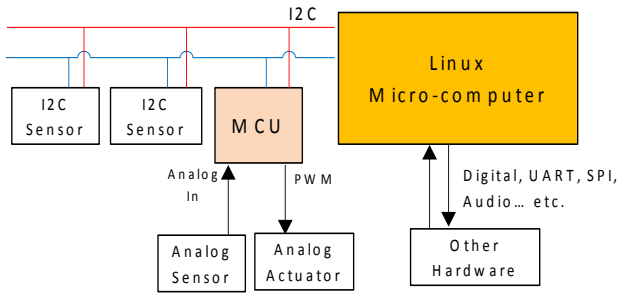

Fig. 5.  I$^2$C configuration.

On the other hand, the UART strategy frees the development process from added I$^2$C addressing management, thus easier to program. A downside is the added complexity in assembly, as there is no longer a single bus to directly connect to, especially if more than a single micro-controller is needed. Another downside, yet minor, is the need to create a software protocol between both peers based on command issuing and data return.

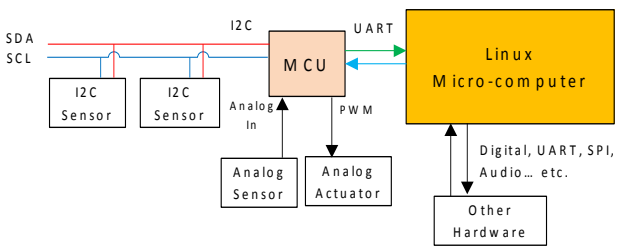

Fig. 6.  UART configuration.

In addition, logic voltage levels must be compatible between both the micro-computer and the micro-controller. In the case of this project's hardware, the Raspberry Pi uses 3.3v logic and the Atmega328 uses 5v. Thus, a voltage level shifter is needed [9]. This is easily implemented via transistor switches. However, this also adds the complexity to the assembly process.

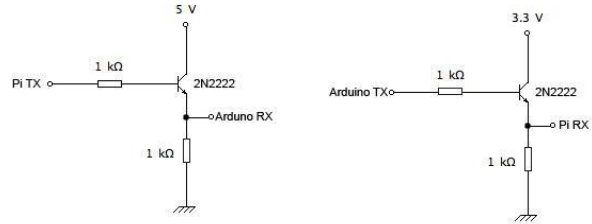

Fig. 7.  Voltage level shifter using transistors.

### 3.3. Obtain and store images from a camera

The main advantage of Linux micro-computers is the fact they operate as desktop computers, but on an embedded systems level. This advantage allows the use of software packages originally targeted for desktop machines. For image capturing as a mission, cameras with a USB interface can be used, especially small-sized webcam (Though due to a USB handling issue on the Raspberry Pi, not all cameras operate correctly).

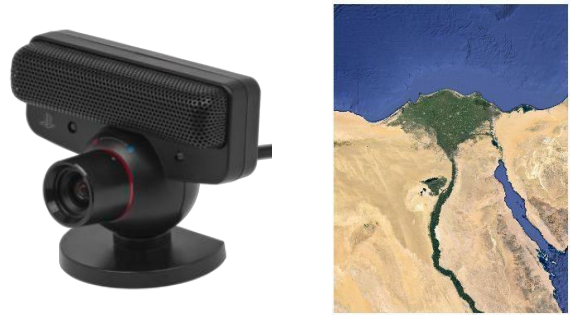

Fig. 8.  Simulated captured image using Sony Playstation Eye camera quality parameters. (Locally developed simulator)

Images from the USB camera can be obtained using code written in Python or C, or using downloadable Linux shell tools. Performance wise, the Linux shell tools proved to be faster during the project's development. Fswebcam [10] is a downloadable Linux tool (Via apt-get or aptitude) that will retrieve the image and store in a given directory on the SD card within ~2 seconds. A Sony Playstation Eye was used for testing.

### 3.4. Receive/Transmit data using a transceiver

In literature, space radio transceivers used on cube-satellites operate from VHF to S band frequency ranges. For computer interface, it was found that usually I$^2$C, SPI, and UART were the methods of interactions with such devices. Due to the high price tags and local unavailability, the project shifted to another type of radio transceivers: Hand-held ham radios, such as the Motorola MJ270R.

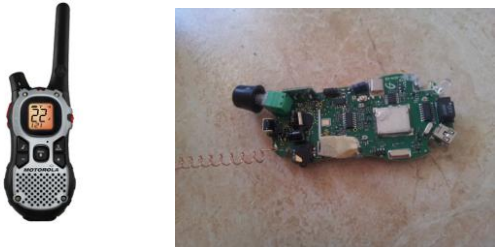

Fig. 9.   Motorola MJ270R unassembled.

However, these hand-held transceivers operate only via sound input through either a microphone or audio AUX. Thus, the use of the micro-computer's audio mechanism as a means of transmission is utilized; sound processing packages and codes were obtained and modified: Robot36 [11] for images and AFSK [12] for binaries. The data is modulated as sound files, transmitted over the transceiver as audio playback of files (with VOX enabled), then demodulated at the ground-station using a locally developed GUI tool.

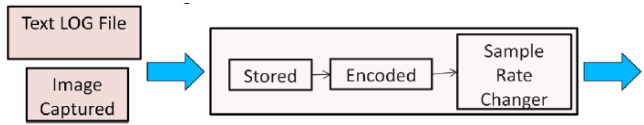

Fig. 10.   Flow of operations of data-to-audio modulation.

But since the Raspberry Pi does not have audio input, the intuitive use of a DTMF tones decoder MT8870 was used. The output is simply a 4-bit digital output that can be read easily using the Raspberry Pi's digital IO pins.

## 4. Android Smartphone

As technological revolution continues in the smartphone industry, it is sensible to manipulate Android devices that come with amusing multi-core processing powers, integrated sensors and camera, and small size on cube-satellites. In fact, it has already been implemented by institutions as NASA and SSTL, with more to come. An onboard computer strategy based on an Android smartphone was also implemented in the project.

The following are the general specifications of most Android smartphones in the market [13] (Eventually, there might be better specifications, depending on the device):

Table 2.   Android smartphone specifications

| Processor | 1 GHz processor (1-4 cores) |
|---|---|
| Memory | 256-1024 Mb RAM<br>Up to 64 Gb SD Card storage |
| Avg. Size | Differs according to device |
| Avg. Weight | 130 gm |
| Power | 3.7v – Current is usage dependant. |
| Sensors | Accelerometer<br>Gyroscope<br>Magnetometer<br>Light-sensor<br>GPS<br>5 MP Camera |
| Interfaces | Micro-USB<br>3.5 mm Audio output<br>Wifi/Bluetooth<br>GSM |

Using Android smartphones have two issues. First, they do not have hardware IO pins as default. Second, they are programmed in Java. The second is more problematic as it limits the options for programming languages, in addition to the fact some libraries are either not available for Java.

The paper will explain the functionalities implemented, as done with Linux micro-computers. How the aforementioned drawbacks are solved are to be discussed as well.

### 4.1.   Computation and operations management

The subroutines management software operates in the same sense as discussed in the micro-computer approach, but with a major addition; a virtual Linux operating system is operated locally on the phone in parallel to the Android operating system. This technique provides major advantages: Utilizing Linux desktop software packages, such as Octave, and the ability to build codes not written in Java. In essence, all the development work done on the micro-computer approach can be ported to the Android phone with minimal adjustments.

The technique is based on an important Linux tool called chroot [14,15]. Briefly, it enables the host operating system to create a shell instance of another operating system using the latter's mounted system files. Because Android is in fact built upon Linux, the use of this utility is possible (Note: It requires system modifications to be enabled. i.e. Rooting). Interacting with the virtual Linux operating system occurs through SSH; a Java SSH library [16] is added to the Android code, which initiates a connection and sends the commands directly. The SD card storage is shared between both peers to facilitate file sharing and usage.

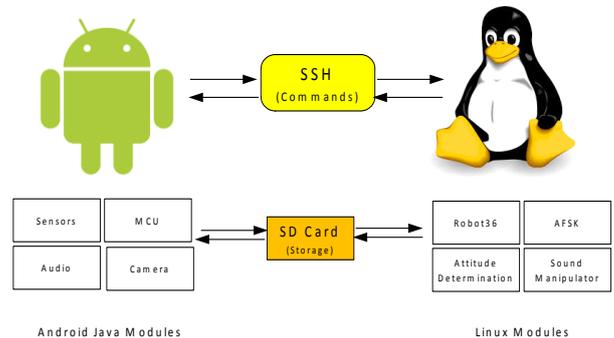

Fig. 11.   Android-Linux interaction.

The output of such strategy is to be demonstrated in specifically in the attitude determination and communications software.

### 4.2   Sensor readings acquisition and actuator control

Since the sensors come already integrated, readings acquisition is done programmatically easy using objects of SensorManager class for sensor readings.

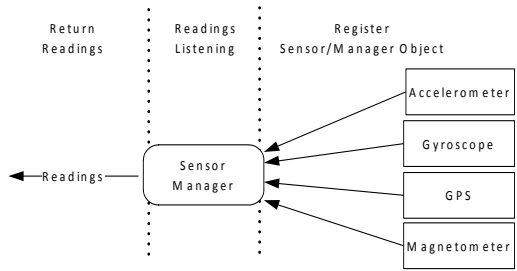

Fig. 12. Sensor manager object operations flow.

However, the problem of no direct hardware IO pins on the smartphone remains, should the system require external interaction with a sensor or actuator. This is solved similarly as in the micro-computer approach: By interfacing with an external micro-controller. This is done through a USB-OTG connector connected from the phone and onto a micro-controller to create a simple UART link [17].

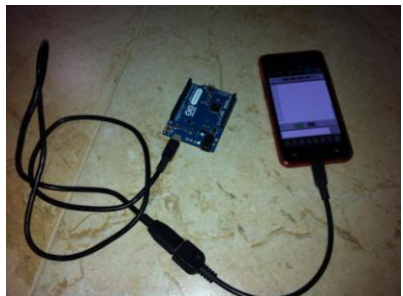

Fig. 13. Android phone connected to an Arduino serially with OTG.

Again, it should be stated that the micro-controllers to be used eventually are Atmega328s, for the purpose of utilizing the Arduino platform and open-source libraries.

After sensor readings are acquired, attitude determination and control commences. Virtual Linux is used to implement the Matlab codes through Octave. This is to reduce development and code porting time, and make it operational on the Android device that by default does not support Matlab coding. First, sensor readings are acquired, sent to Octave via SSH command to the virtual Linux, then the control output is then returned for the phone to control the actuators over the connected micro-controller.

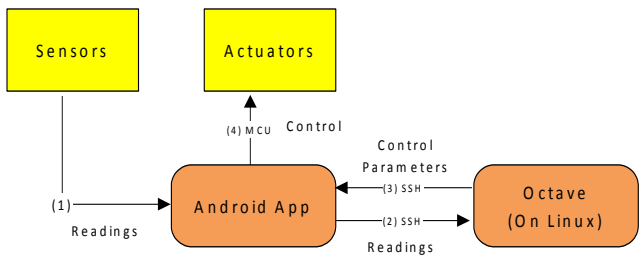

Fig. 14. Workflow of attitude determination and control for Android strategy.

### 4.3 Obtain and store images from a camera

Similarly as the sensors, the camera comes built-in for Android phones. Similarly as well, images are obtained easily using a Java Camera class, but with some modifications to remove user intervention (i.e. Capture images autonomously).

### 4.4. Receive/Transmit data using a transceiver

It was stated that due to the expenses and local unavailability of communications hardware, hand-held radio transceivers were used. It was also stated that these hand-held radios interface only using audio input and output, and that for such reason SSTV and AFSK are used. However, because Android development uses Java for programming, there were no Java audio communications libraries available online, and developing them from scratch would have been time consuming. It is for this reason that virtual Linux is also utilized to operate incompatible modules.

Software modules used on the Linux micro-computer are simply ported to the virtual Linux's local files. Since the files storage of both Android and virtual Linux are shared, the Android application basically signal's the software modules within the virtual Linux to start processing the files through SSH. After the communication sound files are created, they are then audio played-back for the audio transceivers. However, this is fitting only the downlink communications. For uplink communications, the DTMF decoder strategy is to be followed similarly as in the Raspberry Pi approach.

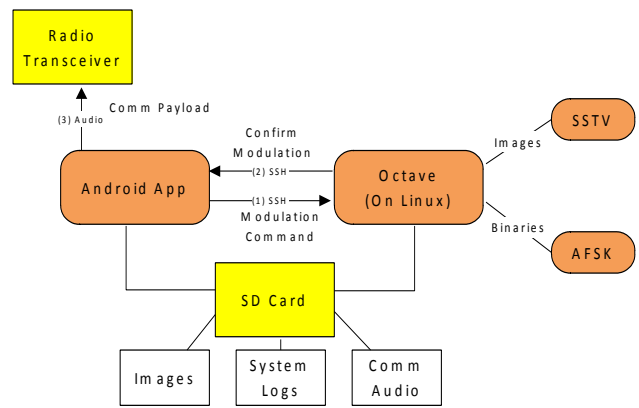

Fig. 14. Workflow of downlink communications for Android strategy.

### 5. Alternatives

Despite the fact functionalities were indeed implemented, it is note-worthy to state that there are currently better and more efficient ways to implement onboard computers using the two strategies. For the Micro-computer approach, a Beaglebone Black [18] can be used instead of the Raspberry Pi; it has more hardware IO pins (Including PWM and A/D converters), stronger processor, smaller form-factor, and it is only 10$ more expensive. This alternative relieves the development from the use of external micro-controllers. However, the Beaglebone does not provide an AUX audio output, thus demanding radio transceivers which operate differently.

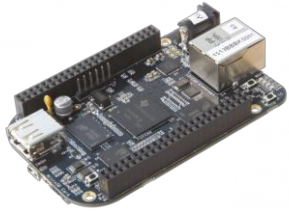

Fig. 15. Beaglebone Black.

As for the Android phone approach, Google has provided a new software development kit to develop Android applications using C programming language, which called the NDK [19,20]. This allows easy porting of applications and packages written in C to the Android phone, instead of operating them remotely on the virtual Linux. Though porting C code might not be a straight forward task and might require meticulous attention to C library dependencies and platform specifics.

## 6. Conclusion

As stated, one of the aims of the B.Sc. project was to implement a computing system with the aid of low-cost off-the-shelf computers. Despite the fact this is not the usual way of development, but the Linux micro-computer and Android phone prove the following:

- They provide needed computing and control functionalities.
- Desktop packages can be used on embedded systems' level.
- Computing can tend to be centralized, rather than distributed.
- Functionalities can still be fulfilled, despite interface issues.

Development of onboard computing systems is heading towards the use of off-the-shelf consumer systems, as demonstrated by top institutions as NASA and SSTL. Even if reliability issues arise, the capacity building resulted from implementing computing systems with space functionalities in mind is worth the efforts. Adding the fact that relatively ultra low-cost systems help attract more and more institutions to build expertise and enter the space race. Elementary and high schools are eligible to start space systems development with such welcoming approaches.

## Acknowledgments

The development team of the B.Sc. project "Subsystems for 2nd Iteration Cairo University Cube-Satellite" greatly appreciates the Nano-Satellite Symposium Office's acceptance of this paper for presentation at Japan in front of a global audience of top engineers.

The team would like to thank the junior team who volunteered to participate and assist in the project, as preparation to develop the next iteration of the cube-satellite. Their efforts are greatly valued.

The team appreciates the patience and acceptance of supervising professors and families for their continued support and assistance during the development of the project.